\documentclass[12pt,thmsa]{article}
\usepackage{sw20lart}

\input tcilatex
\QQQ{Language}{
American English
}

\begin{document}

\author{I. Radinschi\thanks{%
iradinsc@phys.tuiasi.ro} \and Department of Physics, ``Gh. Asachi''
Technical University, \and Iasi, 6600, Romania}
\title{Energy Associated with a Charged Regular Black Hole}
\maketitle

\begin{abstract}
We show that using an adequate coordinate transformation the charged regular
black hole solution given by Ay\'{o}n-Beato and Garcia can be put in the
Kerr-Schild form. Then we use this metric in Kerr-Schild Cartesian
coordinates with a result given by Virbhadra and obtain the energy
distribution associated with this.

Keywords: energy distribution, charged regular black hole

PACS: 04.20.Dw, 04.70.Bw
\end{abstract}

\section{INTRODUCTION}

A problem which still remains unsolved in general relativity is the
energy-momentum localization.

Although an adequate coordinate-independent expression for energy and
momentum density has not given yet, various energy-momentum complexes
including those of Einstein [1]-[2], Landau and Lifshitz [3], Papapetrou
[4], Bergmann [5], Weinberg [6] and M\o ller [7] lead to acceptable results
for many space-times. Cooperstock [8] gave his opinion that the energy and
momentum are confined to the regions of non-vanishing energy-momentum tensor
of the matter and all non-gravitational fields. Also, Chang, Nester and Chen
[9] showed that the energy-momentum complexes are actually quasilocal and
legitimate expressions for the energy-momentum. The above energy-momentum
complexes, except that of M\o ller, need to carry out the calculations in
``Cartesian coordinates''. The results obtained for some well-known
space-times are encouraging [10]-[25].

Aguirregabiria, Chamorro and Virbhadra [15] obtained that several
energy-momentum complexes ``coincide'' for any metric of the Kerr-Schild
class. In [19] Virbhadra established that several energy-momentum complexes
(ELLPW) comply with the quasi-local mass definition of Penrose for a general
non-static spherically symmetric metric of the Kerr-Schild class. He
obtained the expression of the energy distribution for this general metric.

In this paper we first show that using an adequate coordinate transformation
we can express the charged regular black hole (Ay\'{o}n-Beato and Garcia,
(ABG)) [27] metric in Kerr-Schild Cartesian coordinates. Then we compute the
energy distribution in this space-time using the energy expression given by
Virbhadra [19]. The result is the same as we obtained in the Einstein
prescription [26] using the Schwarzschild Cartesian coordinates. We use the
geometrized units $(G=1,c=1)$ and follow the convention that Latin indices
run from $0$ to $3$.

\section{ENERGY\ OF\ THE\ CHARGED\ REGULAR\ BLACK\ HOLE}

A solution to the coupled system of the Einstein field and equations of the
nonlinear electrodynamics was recently given by E. Ay\'{o}n-Beato and A.
Garcia (ABG) [27]. This is a singularity-free black hole solution with mass $%
M$ and electric charge $q$. Also, the metric at large distances behaves as
the Reissner-Nordstr\"{o}m solution. The usual singularity of the RN
solution, at $r=0$, has been smoothed out and now it simply corresponds to
the origin of the spherical coordinates. This solution is given by the line
element

\begin{equation}
ds^2=A(r)dt^2-B(r)dr^2-r^2(d\theta ^2+\sin ^2\theta d\varphi ^2),
\label{1.1}
\end{equation}

where

\begin{equation}
A(r)=B^{-1}(r)=1-\frac{2M}r(1-\tanh (\frac{q^2}{2Mr})).  \label{1.2}
\end{equation}
If the electric charge vanishes we reach the Schwarzschild solution. At
large distances (1) resembles to the Reissner-Nordstr\"{o}m solution and can
be written

\begin{equation}
A(r)=B^{-1}(r)=1-\frac{2M}r+\frac{q^2}{r^2}-\frac{q^6}{12M^2r^4}+O(\frac
1{r^6}).  \label{1.3}
\end{equation}

The Kerr-Schild class space-times have the form

\begin{equation}
g_{ik}=\eta _{ik}-Hl_il_k,  \label{1.4}
\end{equation}

where $\eta _{ik}=diag(1,-1,-1,-1)$ is the Minkowski metric. $H$ represents
the scalar field and $l_i$ is a null, geodesic and shear free vector field
in the Minkowski space-time. We also have

\begin{equation}
\begin{tabular}{c}
$\eta ^{ab}l_al_b=0,$ \\ 
$\eta ^{ab}l_{i,a}l_b=0,$ \\ 
$(l_{a,b}+l_{b,a})l_{\;\,,c}^a\eta ^{bc}-(l_{\;\,,a}^a)^2=0.$%
\end{tabular}
\label{1.5}
\end{equation}

Aguirregabiria, Chamorro and Virbhadra [15] showed that for the space-times
of the Kerr-Schild class the energy-momentum complexes of Einstein, Landau
and Lifshitz, Papapetrou and Weinberg ``coincide''.

We use the transformation

\begin{equation}
u=t+\int A^{-1}(r)dr  \label{1.6}
\end{equation}

and we get

\begin{equation}
dt=du-A^{-1}(r)dr.  \label{1.7}
\end{equation}

Also, we obtain

\begin{equation}
dt^2=du^2+A^{-2}(r)dr^2-2A^{-1}(r)drdu.  \label{1.8}
\end{equation}

Using (8) the line element (1) becomes

\begin{equation}
ds^2=A(r)du^2-2dudr-r^2(d\theta ^2+\sin ^2\theta d\varphi ^2),  \label{1.9}
\end{equation}

that is the static case of the general non-static spherically symmetric
space-time of the Kerr-Schild class used by Virbhadra [19] to calculate the
energy distribution with the energy-momentum complexes of Einstein, Landau
and Lifshitz, Papapetrou and Weinberg (ELLPW).

Now, with the transformations

\begin{equation}
\begin{tabular}{c}
$T=u-r,$ \\ 
$x=r\sin \theta \cos \varphi ,$ \\ 
$y=r\sin \theta \sin \varphi ,$ \\ 
$z=r\cos \theta ,$%
\end{tabular}
\label{1.10}
\end{equation}

the metric given by (9) can be written

\begin{equation}
ds^2=dT^2-dx^2-dy^2-dz^2-(1-A)\times [dT+\frac{xdx+ydy+zdz}r]^2.
\label{1.11}
\end{equation}

For (11) we have $H=1-A$ and $l_i=(1,\frac xr,\frac yr,\frac zr)$. We
evaluate the energy using the expression obtained by Virbhadra [19], (see
Eq. (31) therein), in the case of a general non-static spherically symmetric
space-time of the Kerr-Schild class, and which is the same in the (ELLPW)
prescriptions. Our case is the static case, which is a special case of the
metric considered by Virbhadra. We get

\begin{equation}
E(r)=\frac r2(1-A(r)).  \label{1.12}
\end{equation}

We obtain for the energy distribution of the ABG black hole

\begin{equation}
E(r)=M(1-\tanh (\frac{q^2}{2Mr}))  \label{1.13}
\end{equation}

and

\begin{equation}
E(r)=M-\frac{q^2}{2r}+\frac{q^6}{24r^3M^2}-\frac{q^{10}}{240M^4r^5}+O(\frac
1{r^6}).  \label{1.14}
\end{equation}

Also, for (14) we can write

\begin{equation}
E(r)=E_{RN}(r)+\frac{q^6}{24M^2r^3}-\frac{q^{10}}{240M^4r^5}+O(\frac 1{r^6}),
\label{1.15}
\end{equation}

where the term $E_{RN}(r)$ represents the energy of the
Reissner-Nordstr\"{o}m solution that corresponds to the Penrose [28]
quasi-local mass definition.

We define $E^{^{\prime }}=\frac{E(r)}M$, $Q=\frac qM$ and $R=\frac rM$ and
we have $E^{^{\prime }}=1-\tanh (\frac{Q^2}{2R})$. We plot the expression of 
$E^{^{\prime }}$ in the Figure 1 ($E^{^{\prime }}$ on Y-axis is plotted
against $R$ on X-axis, for $Q=0.1,...,1$).

\section{DISCUSSION}

Bondi [29] gave his opinion that a nonlocalizable form of energy is not
admissible in relativity.

For the charged regular black hole solution given by Ay\'{o}n-Beato and
Garcia we make a coordinate transformation that allow us to evaluate the
energy distribution using the expression of the energy obtained by Virbhadra
[19] (see Eq. (31) therein), for a general non-static spherically symmetric
space-time of the Kerr-Schild class. The new form of the (ABG) metric is a
special static case of the metric considered by Virbhadra [19]. The energy
distribution depends on the mass $M$ of the black hole and electric charge $%
q $. Also, the result is the same as we obtained [26] in the Einstein
prescription using the Schwarzschild Cartesian coordinates. Our result
sustain the viewpoint of Virbhadra [19] that the Einstein energy-momentum
complex is the most adequate to evaluate the energy distribution of a given
space-time.

\textbf{Acknowledgments}

I am grateful to Professor K. S. Virbhadra for his helpful advice.

\end{document}